\documentclass[aps,prl,superscriptaddress,reprint]{revtex4-2}

\usepackage{graphicx,xcolor,siunitx,amssymb,amsmath}
\usepackage[version=4]{mhchem}
\usepackage{hyperref}
\usepackage[columnwise]{lineno} 

\newcommand{\fp}[1]{\textbf{\textsf{#1}}}
\newcommand{\figref}[1]{Fig.\,\ref{#1}}

\newcommand{\figrefp}[2]{Fig.\,\ref{#1}\,\fp{#2}}
\newcommand{\Figrefp}[2]{Figure\,\ref{#1}\,\fp{#2}}

\newcommand{\figtitle}[1]{\textbf{\textsf{#1}}}

\newcommand{\paratitle}[1]{
\noindent \textbf{#1}}

\newcommand{\diff}[2]{\frac{\mathrm{d}{#1}}{\mathrm{d}{#2}}}

\begin{document}

\title{Transient Chirality in the Gelation of Adhesive Spinner Monolayers}

\author{Yujie Jiang}
\email{yjjiang@ustb.edu.cn}
\affiliation{School of Mathematics and Physics, University of Science and Technology Beijing, Beijing 100083, China}
\author{Haiquan Li}
\affiliation{School of Mechanical Engineering, University of Science and Technology Beijing, Beijing 100083, China}
\author{Yiting Liu}
\affiliation{21C LAB, Contemporary Amperex Technology Company, Limited, Ningde 352000, China}
\author{Haoran Li}
\affiliation{21C LAB, Contemporary Amperex Technology Company, Limited, Ningde 352000, China}
\author{Yang Cui}
\affiliation{21C LAB, Contemporary Amperex Technology Company, Limited, Ningde 352000, China}

\date{\today}

\begin{abstract}

Active systems of self-rotating elements inherently exhibit chirality, making them of fundamental interest due to parity violation. 
Using large-scale hydrodynamic simulations, we investigate the gelation of adhesive spinners confined to quasi-2D monolayers at low Reynolds numbers.
Unlike the coarsening dynamics of passive colloids, spinner gelation follows a different pathway, displaying structural chirality during the early stages of aggregation.
However, this chirality dissipates upon dynamical arrest, resulting in a final gel structure that resembles a conventional colloidal gel. 
As a result, we find no sign of odd mechanical responses.
Nonetheless, the elastic modulus and gelation time remain tunable through spinning activity, providing a new avenue for the bottom-up design of programmable soft materials.

\end{abstract}

\maketitle

\section*{Introduction}

Soft particulate gels are ubiquitous in industrial applications, ranging from daily consumer products to pharmaceuticals and biotechnology\,\cite{Nature.453.499,NatPhys.19.1178,JRheol.69.35}.
These gels usually form when attractive colloidal particles undergo Brownian motion, aggregate into open clusters, and eventually percolate throughout the system, i.e. the colloidal gel\,\cite{SciAdv.5.eaav6090,NatCommun.11.3558}.
The resulting porous network imparts unique mechanical and rheological properties\,\cite{RevModPhys.89.035005, PhysRevLett.103.208301}.
While non-equilibrium protocols, such as thermal annealing\,\cite{NatPhys.20.1171} and external flow\,\cite{SoftMatter.11.4640}, have been developed to tune gel structures, recent studies combine active matter and particulate gels to achieve programmable properties\,\cite{PhysRevLett.119.058001,ACSNano.13.560,NatPhys.19.1680,PhysRevLett.133.228301}.
For instance, self-propelled active particles have been utilized to regulate mesoscale dynamics and, consequently, the structure within gels through local energy injection\,\cite{PNAS.121.e2407424121}.


Beyond directional swimmers, spinners -- particles driven by an active torque -- also constitute an important class of active matter\,\cite{PhysRevLett.112.075701,NatPhys.21.146,EPL.139.67001}.
These systems typically operate in hydrodynamic environments, which, in the absence of Brownian diffusion, enable translational motion at a collective level\,\cite{Nature.607.287}.
The inherent chirality of rotational dynamics gives rise to diverse self-assembly behaviors and collective motions\,\cite{NatPhys.17.1260,NatPhys.19.1680,NatCommun.6.5994}.
Moreover, the resulting non-reciprocal interactions can lead to unique, parity-violating odd responses, such as odd viscosity\,\cite{AnnuRevCondensMatterPhys.14.471} and Hall-like effect\,\cite{PNAS.119.e2201279119}.
While extensive research has been conducted on \textit{chiral active fluids}\,\cite{NatPhys.18.154,PhysRevLett.130.158201,NatPhys.17.1260}, the interplay between rotational activity and gelation remains less explored, despite its scientific significance as highlighted in recent works\,\cite{NatPhys.19.1680,PhysRevRes.6.023186}.


In this study, we employ large-scale hydrodynamic simulations to investigate the gelation of adhesive spinners under inertialess conditions.
The system is confined to quasi-2D (i.e., monolayers) to accentuate chiral effects.
Our simulation scheme couples the Lattice Boltzmann Method (LBM) with the Discrete Element Method (DEM)\,\cite{ComputFluids.189.1}, incorporating lubrication corrections to fully resolve fluid-solid interactions.
Conventional colloidal gelation proceeds via arrested phase separation, where spinodal decomposition textures are dynamically arrested at a percolating state\,\cite{JPhysCondensMatter.19.323101}.
Unlike passive colloids, spinning rotors do not undergo diffusive motion, making it intriguing to examine the emergence of chirality and its potential contribution to odd responses.
We find that adhesive spinners gel through a different route, yet end up with a similar structure to colloidal gels, irrespective of spinning activity.
Structural chirality appears locally during early-stage clustering, but dissipates as clusters grow into a percolating network.
Thus, self-rotation-induced chirality is not retained in the final gel, which presents isotropic structure and rheology.
However, the elastic modulus and gelation rate vary as functions of spinning activity, suggesting a novel approach to tailoring particulate gels.

\section*{Results}

\begin{figure*}[htbp]
\centering
\includegraphics[width=\textwidth]{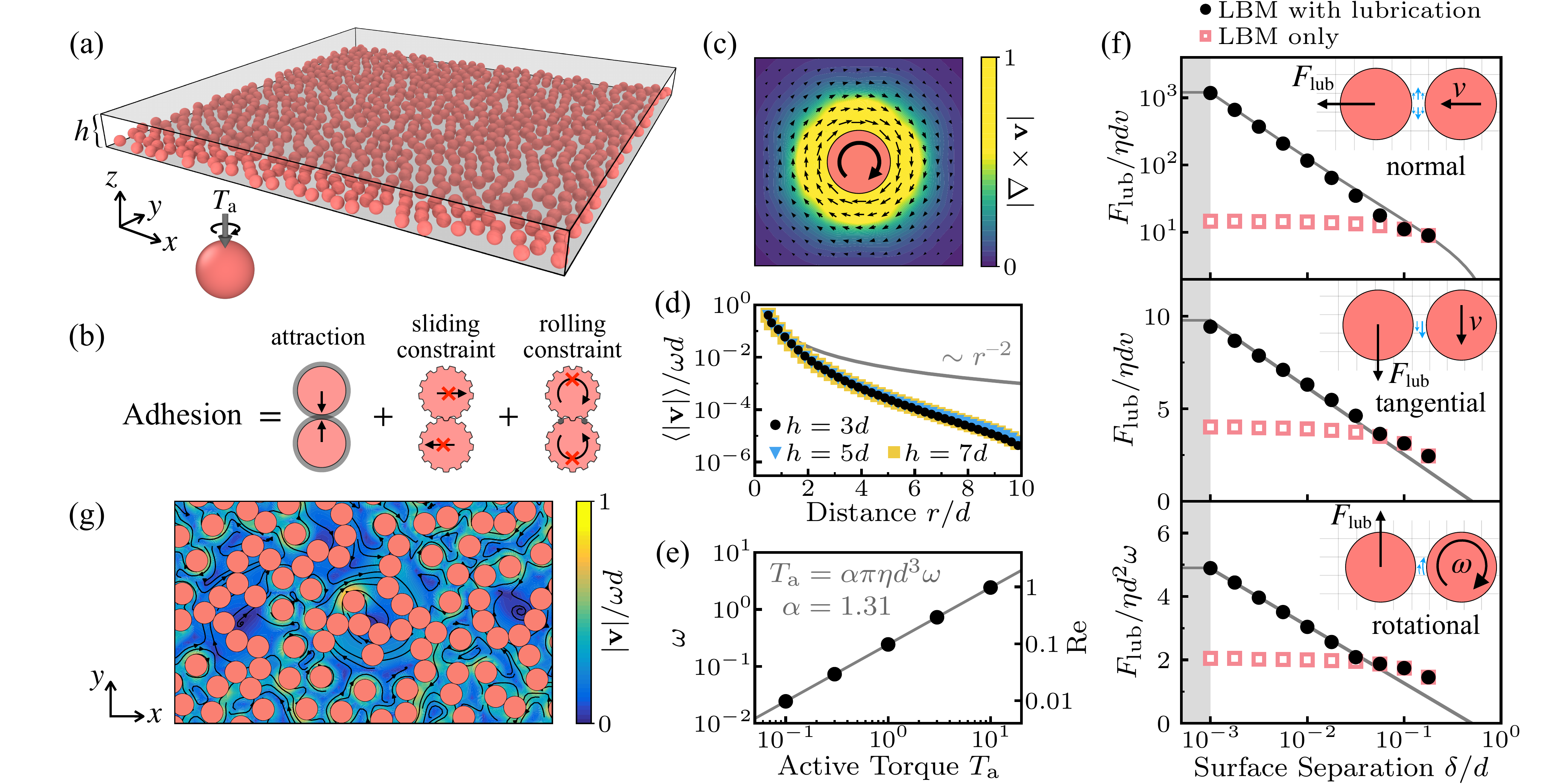}
\caption{%
\figtitle{Simulation setup.}
\fp{a}~Simulation setup of the quasi-2D spinner monolayer.  
\fp{b}~Adhesion model depicting the central attractive force and tangential constraints, including sliding and rolling resistance.
\fp{c}~Flow field in the $x$-$y$ plane at $z=0.5d$, showing velocity vectors (arrows) and vorticity (colormap) around an isolated spinner.
\fp{d}~Normalized velocity profiles as a function of radial distance $r/d$ at different box heights $h$. The solid gray line represents $\langle | \mathbf{v} | \rangle/\omega d = 0.125r^{-2}$.
\fp{e}~Angular velocity $\omega$ as a function of active torque $T_{\text{a}}$. Linear fitting (solid gray line) yields $T_{\text{a}} = \alpha \pi \eta d^3 \omega$ with $\alpha=1.31$.
\fp{f}~Comparison of lubrication forces $F_{\mathrm{lub}}$ between simulations with and without lubrication corrections. Top: normal approach; middle: tangential sliding; bottom: rotation. Solid gray lines denote theoretical predictions\,\cite{Suspension}. The shaded region ($\delta < 10^{-3}d$) indicates the inner cutoff of the lubrication model.
\fp{g}~Snapshot of a $\phi=0.4$ spinner monolayer during gelation. Curved arrows indicate flow streamlines, while the color map represents velocity magnitude.
}
\label{fig1}
\end{figure*}

\paratitle{System characterization.}
Our system comprises $N$ spherical spinners (of diameter $d$) suspended in a Newtonian fluid (of viscosity $\eta$ and density $\rho$).
These particles are confined within a thin square simulation box with lateral dimensions $L_{x,y} = 120d$ and height $h=3d$ along the $z$-axis, \figrefp{fig1}{a}.
Each spinner experiences an active torque $T_{\text{a}}$ applied along the $-z$ direction, inducing clockwise self-rotation.
While periodic boundaries are applied to the $x$- and $y$-directions, we introduce two flat walls at $z=0$ and $z=h$, and confine the particles to the bottom monolayer at $z=0.5d$.
This setup mimics experimental conditions where density-mismatched particles, such as the hematite beads\,\cite{CommunPhys.7.291,PhysRevRes.3.L042021}, sediment onto a substrate.
For simplicity, we consider isotropic adhesion instead of complex interactions (such as magnetic-dipolar forces), allowing for a more fundamental scope.
The adhesion applies within a short range ($\zeta_{\text{a}} = 0.01d$), consisting of central attraction, tangential friction and rolling resistance, \figrefp{fig1}{b}.
All the three components are depicted by modified Hookean models with a unified spring constant $k$, which is sufficiently large ($U_{\mathrm{adh}} \equiv \frac{1}{2}k\zeta_{\text{a}}^2 \gg T_{\text{a}}$) to ensure strong adhesion.
As a result, relative motions between adhered particles is effectively constrained.


The dynamics of the adhesive spinners are implemented in DEM using LAMMPS\,\cite{CompPhysComm.271.108171}, while hydrodynamic interactions are captured using LBM.
To balance efficiency and accuracy, we set the LBM lattice spacing to $\Delta x = 0.25d$, which provides minimal yet sufficient resolution to reproduce the Stokes drag on a single-particle level.
Within this resolution, lubrication corrections are applied between particle pairs without compromising accuracy, \figrefp{fig1}{f}.
By setting an inner cutoff $10^{-3}d$ to prevent divergence at contact, this simulation scheme well captures both near- and far-field hydrodynamics.
More simulation details can be found in the Methods section.


The flow field surrounding an isolated spinner, \figrefp{fig1}{c}, exhibits a radial decay.
\Figrefp{fig1}{d} shows that the decay of the averaged velocity $\langle|\mathbf{v}|\rangle$ is faster than an expected inverse square law\,\cite{Suspension}, likely due to additional hydrodynamic resistance from the no-slip walls (particularly the bottom one).
In this work, we use a thin simulation box with a height of $h=3d$, which yields velocity profiles similar to those obtained with larger $h$, \figrefp{fig1}{d}.
The rotation speed $\omega$ is proportional to the applied torque $T_{\text{a}}$, and a linear fit to $T_{\text{a}}=T_{\text{a}}(\omega)$ reveals a shift factor of $\alpha = 1.31$ from the Stokes law, \figrefp{fig1}{e}.
This deviation also arises from the no-slip bottom wall.


To manifest the role of chirality in gelation, we systematically eliminate potential confounding factors.
Our spinners are athermal so that all motion is solely caused by self-rotation.
The Reynolds number, defined as $\text{Re} \equiv \omega \rho d^2/\eta = 0.1$, is sufficiently low to suppress inertial effects such as levitation\,\cite{PhysRevLett.125.228002} and secondary flows\,\cite{PhysRevLett.130.188202}.
Hydrodynamic repulsion due to the Magnus effect\,\cite{NatCommun.6.5994} is also negligible.
The area fraction $\phi \equiv N\pi d^2 /4L_{x}L_{y}$ is mainly fixed at $0.4$, while we also probe the system from $\phi=0.3$ to $0.5$.
Simulations are initialized with a random configuration without overlap, and evolve for $50\times2\pi/\omega$ (i.e., 50 laps) until reaching a steady state.


\begin{figure*}[htbp]
\centering
\includegraphics[width=\textwidth]{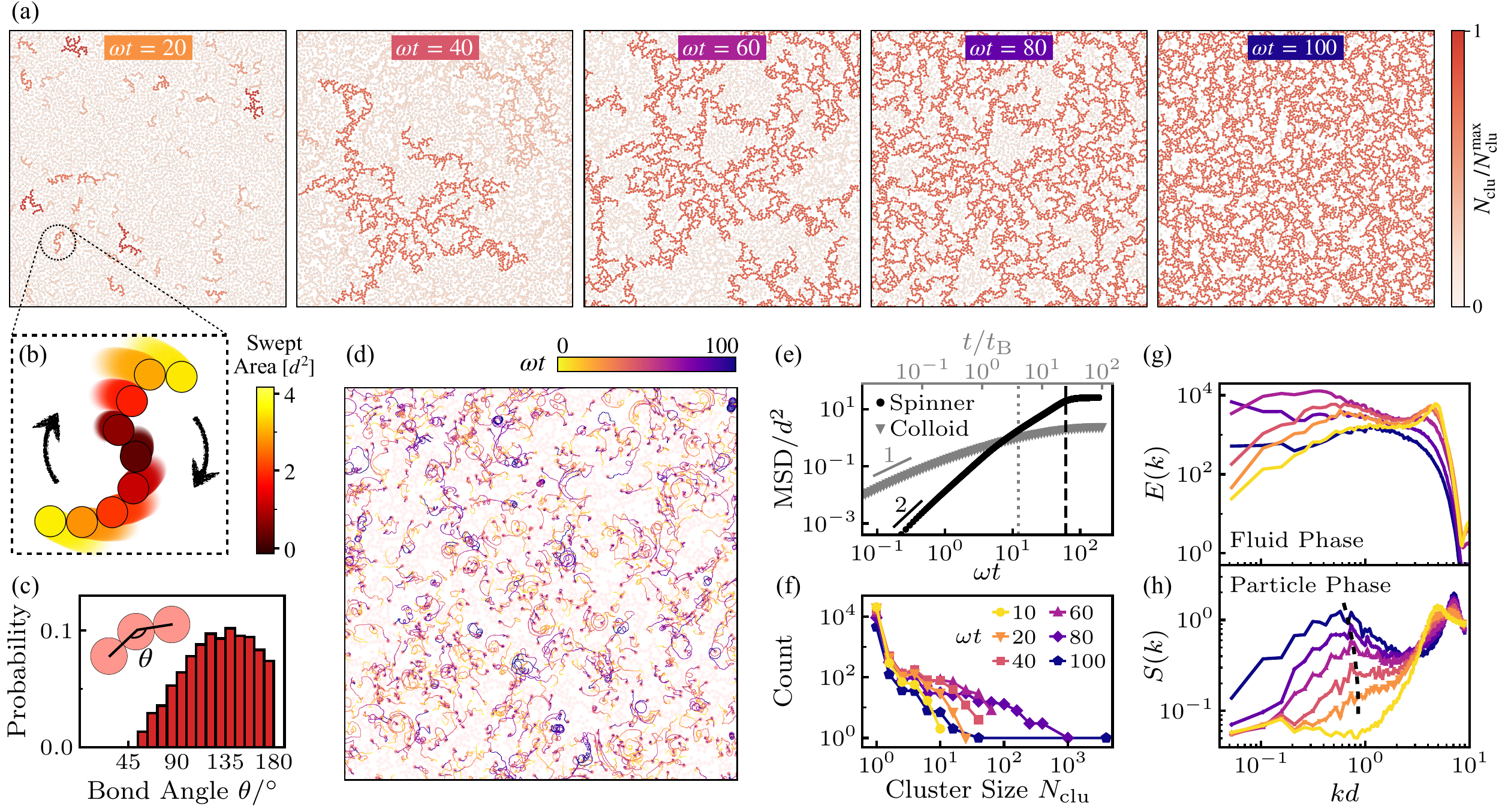}
\caption{%
\figtitle{Spinner gelation.}
\fp{a}~Snapshots of spinner gelation at $\phi=0.4$ and $\mathrm{Re} = 0.1$ (Supplementary Video S1). Color indicates normalized cluster size $N_{\text{clu}}/N_{\text{clu}}^{\text{max}}$.
\fp{b}~Isolated particles preferably bond to the outer regions of these clusters due to the larger swept area, promoting the growth of S-shaped structures.
\fp{c}~Distribution of bond angles $\theta$ at $\omega t = 20$.
\fp{d}~Trajectories of randomly selected spinners during gelation. Color indicates the progression of $\omega t$.
\fp{e}~Mean squared displacement (MSD) during colloidal and spinner gelation. Time is normalized by the Brownian time $t_{\mathrm{B}}\equiv \pi\eta d^3/2k_{\mathrm{B}}T$ for colloids and by the spinning frequency $\omega$ for spinners. Solid lines indicate slopes of 2 (ballistic) and 1 (diffusive), respectively. Dashed vertical lines denote percolation points.
\fp{f}~Evolution of the cluster size distribution during spinner gelation. The same color scheme is used in (g) and (h).
\fp{g}~Evolution of the energy spectrum $E(k)$ of the flow fields in the $x$-$y$ plane ($z=0.5d$), considering only the fluid phase. 
\fp{h}~Evolution of the structure factor $S(k)$, computed from the particle configuration. The dashed black line indicates the peak position as a visual guide.
}
\label{fig2}
\end{figure*}

\medskip 
\paratitle{Spinner gelation.}
As particles spin and agitate the surrounding fluid, the consequent flow in turn drives spinners to move translationally and aggregate under adhesive interactions, \figrefp{fig1}{g}.
For small clusters, the summed active torque causes them to rotate collectively around their center of mass, while adhesion restricts internal relative motion.
More spinners then adhere to the periphery of these clusters, forming chiral S-shaped clusters, \figrefp{fig2}{a} (left, $\omega t = 20$).
Due to clockwise rotation, particles preferentially bond at the leading edge of the rotating clusters, where the swept area is larger than the interior, \figrefp{fig2}{b}.
Statistics on bond angles reveals a peak at $\theta\approx\SI{140}{\degree}$, \figrefp{fig2}{c}.
This gives a characteristic cluster size of $\sim 10$ particles, beyond which inward growth may occur.


Apart from structural effects, chirality also manifests in the dynamics.
We randomly select \num{1000} particles, most of which present circular trajectories in a clockwise direction, \figrefp{fig2}{d}.
Unlike typical chiral active fluids\,\cite{NatCommun.9.931,PhysRevRes.2.013358}, these trajectories do not exhibit significant diffusive behavior.
This is consistent with the mean squared displacement (MSD) in \Figrefp{fig2}{e} (black), which indicates that spinner motion remains predominantly ballistic.


As aggregation proceeds, the growth of chiral clusters slows down.
For a cluster of $N_{\mathrm{clu}}$ spinners, it rotates around its center of mass with a total active torque $N_{\mathrm{clu}}T_{\mathrm{a}}$.
Since the cluster is neither chain-like nor densely packed, its spatial size scales as $l_{\mathrm{clu}} \propto {N_{\mathrm{clu}}}^\alpha$ with $0.5 < \alpha < 1$.
Assuming the rotational drag coefficient $c_{\mathrm{r}}$ follows the cubic relation ($c_{\mathrm{r}} \propto {l_{\mathrm{clu}}}^3$, see Supplementary Note 1), the cluster’s rotational speed is estimated as:
\begin{equation}
\omega_{\mathrm{clu}} \propto \frac{N_{\mathrm{clu}}T_{\mathrm{a}}}{c_{\mathrm{r}}} \sim {N_{\mathrm{clu}}}^{1 - 3\alpha}~~~(0.5< \alpha < 1).\label{eq1}
\end{equation}
It is obvious that the rotational speed decreases rapidly as clusters grow larger.
Additionally, hydrodynamic resistance from the solid walls further slows their rotation.


As the rotation of S-shaped clusters slows, their structural chirality becomes less pronounced.
On the one hand, differences in cluster spinning speeds (Eq.\,\eqref{eq1}) induce rotational decoherence.
On the other hand, while individual S-shaped clusters are chiral, subsequent cluster-cluster aggregation randomizes their spatial orientations.
The growth history of clusters is shown in \Figrefp{fig2}{a}.
At $\omega t = 40$, transient chirality largely vanishes as clusters merge into larger, open structures.
By $\omega t = 60$, the largest cluster grows rapidly connects with other clusters, percolating the system.
Visually, the resulting isotropic network lacks global handedness.


For better comparison, we also perform Langevin dynamics to simulate passive colloidal gels with strong adhesion ($U_{\mathrm{adh}} \gg k_{\mathrm{B}}T$).
Although both systems percolate, their gelation pathways differ.
In colloidal gels, particles diffuse until forming a space-spanning network, \figrefp{fig2}{e} (gray), upon which coarsening proceeds slowly (Supplementary Video S2).
Conversely, spinner motion is initially ballistic due to convective flow and transitions to super-diffusion before becoming arrested in a percolating state, \figrefp{fig2}{e} (black).
Moreover, clusters in colloidal gels grow uniformly (Supplementary Note 2), with all particles incorporating into the final gel network over time.
By contrast, even after percolation, a fraction of isolated monomers remains in the spinner gel, \figrefp{fig2}{f}.
As illustrated in \Figrefp{fig2}{a}, adhesive spinners tend to percolate first and subsequently undergo internal coarsening, resembling ``viscoelastic phase separation gel'' formation\,\cite{SciAdv.6.eabb8107}, yet independent of $\phi$ (Supplementary Note 3).
The largest cluster is quite loose at the percolation point (\figrefp{fig2}{a}, $\omega t = 60$), with monomers and small clusters trapped inside enclosed loops.
Beyond this stage, no visible chirality is retained at large scales, and local configurations become increasingly compact over time.


Colloidal gelation proceeds as an arrested phase separation\,\cite{JPhysCondensMatter.19.323101}, where a characteristic lengthscale emerges and grows and gradually stabilizes over time.
In the structure factor of spinner gel, nevertheless, a time-invariant lengthscale $\xi\equiv 2\pi/k_{\mathrm{peak}}$ arises from homogeneity and become increasingly significant (\figrefp{fig2}{h}) instead of a growing process in typical phase separation (Supplementary Note 2).
In contrast with the particle configuration, the energy spectrum of the flow field in the $x$-$y$ plane indicates a transition in lengthscale, \figrefp{fig2}{g}.
Initially, localized flow fields form around individual spinners.
As aggregation progresses, an inverse energy cascade occurs, characterized by a peak shift in $E(k)$ toward lower wavenumbers $k$, suggesting the emergence of collective flow driven by cluster rotations.
While kinetic energy at large scale becomes increasingly significant, a sudden decay is observed upon percolation, which greatly arrests the motion of both particles (\figrefp{fig2}{e}) and fluid (\figrefp{fig2}{g}).

\begin{figure}[tbp]
\centering
\includegraphics[width=8.6cm]{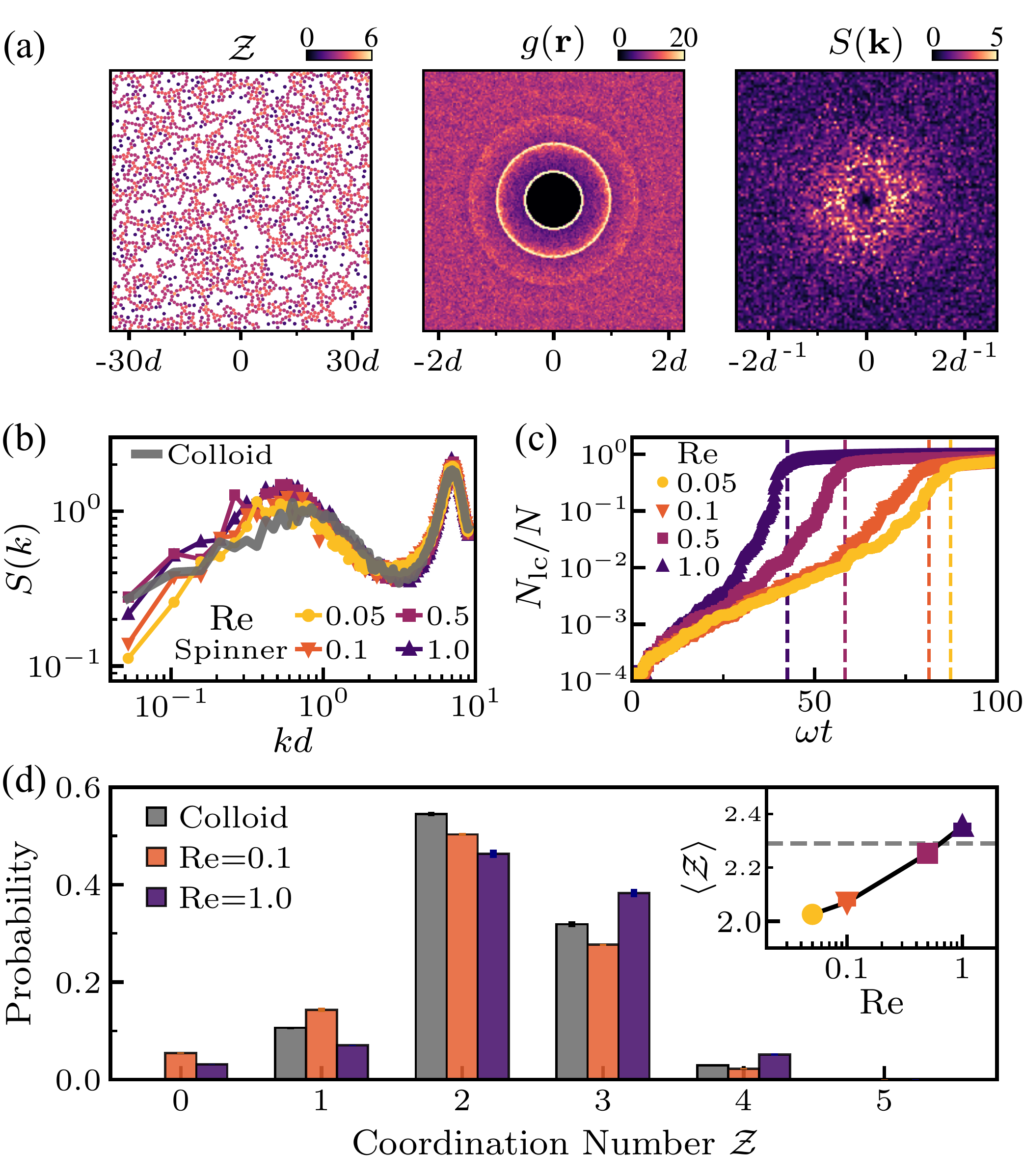}
\caption{%
\figtitle{Gel structure and dynamics.}
\fp{a}~Structure of a spinner gel at $\phi=0.4$ and $\mathrm{Re} = 0.1$. Left: particle configuration; middle: pair distribution function; right: static structure factor.
\fp{b}~Radial-averaged structure factors $S(k)$ for a colloidal gel and spinner gels at $\phi=0.4$.  
\fp{c}~Evolution of the fraction of particles in the largest cluster $N_{\mathrm{lc}}/N$, for different Re. Dashed lines denote percolation points.
\fp{d}~Distribution of coordination number $\mathcal{Z}$. Inset: mean coordination number $\langle \mathcal{Z} \rangle$ as a function of Re. The gray dashed line refers to the colloidal gel.
}
\label{fig3}
\end{figure}

\medskip 
\paratitle{Structural analysis.}
Remarkably, despite of their different gelation routes, the final structures of colloidal and spinner gels are quite similar.
While the spinner gel configuration appears similarly heterogeneous, multi-scaled, and achiral, both the pair distribution function $g(\mathbf{r})$ and structure factor $S(\mathbf{k})$ are isotropic without evident angular dependence, \figrefp{fig3}{a}.
Quantitatively, the structural differences between the two gels are subtle, and their characteristic lengthscales are comparable, \figrefp{fig3}{b}.
The fractal dimension at intermediate scales seems to be lower in spinner gels (as indicated by the shallower slope in $S(k)$, \figrefp{fig3}{b}), likely due to the presence of chain-like strands, \figrefp{fig3}{a} (left).
The distribution of coordination number $\mathcal{Z}$ is also similar between the two systems, \figrefp{fig3}{d}.
The main difference lies in the presence of isolated monomers ($\mathcal{Z}=0$), which are absent in colloidal gels.


Interestingly, the spinning activity $T_{\mathrm{a}}$, or equivalently the Reynolds number Re, plays little role on the final structure.
Within the inertialess regime ($0.05 \leq \mathrm{Re} \leq 1.0$), the structure factors are nearly identical as shown in \Figrefp{fig3}{b}, with only a slight difference in the large-scale homogeneity.
At the particle level, an increase in Re leads to a higher fraction of bonded spinners, \figrefp{fig3}{d}.
As activity increases, the number of monomers decreases, while the fraction of particles with coordination number $\mathcal{Z} \geq 3$ (i.e. the branching point\,\cite{NatPhys.19.1178}) increases as shown in \figrefp{fig3}{d}.
In general, higher spinning activity results in a greater average coordination number $\langle \mathcal{Z} \rangle$, \figrefp{fig3}{d} (inset).

Despite the structural similarities, gelation dynamics exhibit a clear dependence on Re.
As spinning becomes faster, clustering and gelation also accelerate.
While this is expected as the flow velocity is, in principle, proportional to the spinning speed $\omega$, normalizing time by $\omega^{-1}$ does not collapse the growth curves of the largest cluster $N_{\mathrm{lc}}/N$, \figrefp{fig3}{c}.
This trend also applies to percolation points (dashed lines), indicating that the acceleration in dynamics exceeds simple linear scaling.
A possible explanation is that, faster rotation drives spinners to better overcome the lubrication barrier before making contact.
This is further supported by the larger coordination number at higher $\mathrm{Re}=1.0$, \figrefp{fig3}{d}.


\medskip 
\paratitle{Absence of odd rheology.}
Transient chirality is observed at various spinning speeds Re and concentrations $\phi$ (Supplementary Note 3), yet none of these chiral structures persist in the final gel, as evidenced by the isotropic $g(\mathbf{r})$ and $S(\mathbf{k})$.
To further confirm the absence of chirality, we measure the shear rheology upon the removal of the active torque $T_{\mathrm{a}}$.
In particular, $T_{\mathrm{a}}$ is turned off after gelation, and the system is allowed to fully relax under overdamped conditions.
Once the gels reach equilibrium, steady shear is imposed in opposite directions separately, and the resulting stress $\sigma$ is measured as a function of strain $\gamma$.

\begin{figure}[tbp]
\centering
\includegraphics[width=8.6cm]{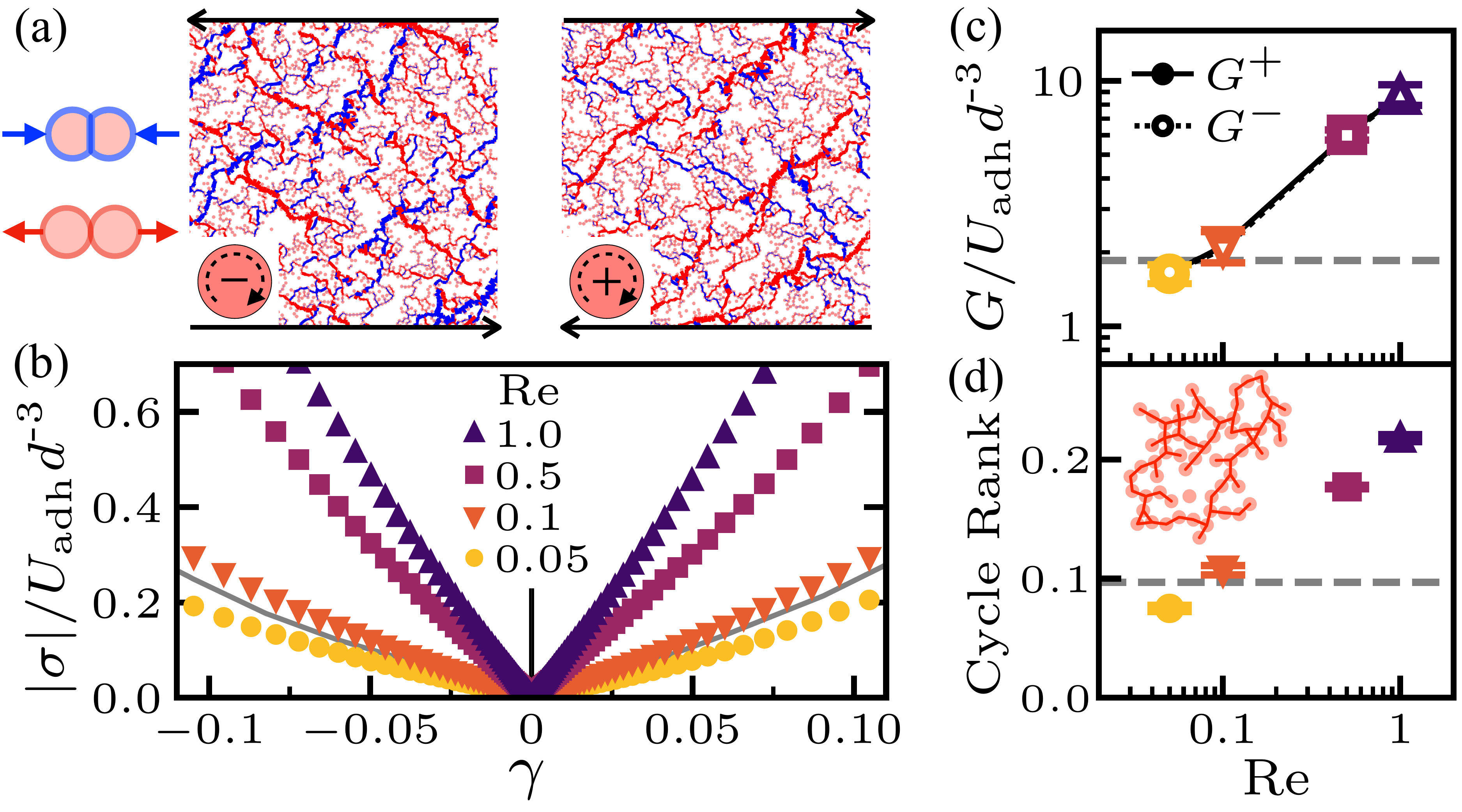}
\caption{%
\figtitle{Rheology and topology.}
\fp{a}~Force chains under shear at $\gamma=-0.1$ (left, misaligned shear) and $\gamma=0.1$ (right, aligned shear). Blue indicates tension, while red represents compression. Bond width is proportional to the force magnitude.
\fp{b}~Stress response under shear applied in opposite directions. The solid gray line represents the colloidal gel rheology.
\fp{c}~Elastic modulus $G$ as a function of spinning activity Re. Solid and open symbols denote aligned ($+$) and misaligned ($-$) shear moduli, respectively. The gray dashed line represents the colloidal gel modulus.
\fp{d}~Cycle rank as a function of Re. The gray dashed line represents the colloidal gel. Inset: schematic nodes (pink disks) and loops (red bonds).
}
\label{fig4}
\end{figure}

Chiral systems, such as spinners, naturally exhibit odd mechanical responses, including odd viscosity and odd elasticity\,\cite{AnnuRevCondensMatterPhys.14.471}.
However, for a spinner gel at $\mathrm{Re} = 0.1$, force transmission remains nearly identical when sheared along different directions, \figrefp{fig4}{a}, though the compressive resistance (blue) appears slightly greater under misaligned ($-$) shear.
Regardless of spinning activity, the absolute values of the responding stress $|\sigma|$ are almost symmetric, as shown in \Figrefp{fig4}{b}.
The elastic moduli, extracted from linear fits to $\sigma=G\gamma$ at small strain $|\gamma| < 0.01$, show no dependence on the shear direction, \figrefp{fig4}{c}.
This is consistent with the isotropic gel structure, \figrefp{fig3}{a}.
Thus, spinner gels are mechanically achiral, as transient chirality from self-rotation is not `memorized' by dynamical arrest.


Although no odd response is observed, the elastic modulus does depend on spinning activity Re, \figrefp{fig4}{c}.
Consistent with the coordination number $\mathcal{Z}$ in \Figrefp{fig3}{d} (inset), the modulus increases with Re in a power law manner with an exponent $\approx 0.5$.
At $\mathrm{Re} = 0.1$, the modulus $G$ is comparable to that of the colloidal gel, even though around \SI{5}{\percent} spinners are isolated (\figrefp{fig3}{d}) and thereby do not contribute to elasticity.
Compared with the colloidal gel, the density of branching points is also lower in the spinner gel at $\mathrm{Re} = 0.1$.
Thus, a softer spinner gel would be expected under this case.
This inconsistency may arise from differences in network topology, which significantly influence gel rheology\,\cite{PNAS.121.e2316394121,ACSNano.18.28622}.
In particular, the cycle rank, denoting the ratio of loops to nodes in a network, accurately captures the modulus variation in our gels, \figrefp{fig4}{d}.
Consistent with polymeric systems\,\cite{Macromolecules.56.9359,SoftMatter.20.7103}, the modulus increases with the cycle rank.
At $\mathrm{Re} = 0.1$, the spinner gel displays a cycle rank similar to that of the colloidal gel.
Although a higher coordination number (and branching density) in the colloidal gel is expected to increase the modulus, floppy modes associated with large loops (Supplementary Note 4) may counteract this stiffening effect, ultimately leading to similar moduli.

\section*{Discussion}

In summary, we investigate the gelation of adhesive rotors in a quasi-2D monolayer, revealing a distinct `top-down' gelation pathway where percolation precedes inward coarsening. 
Despite the intrinsic chirality of self-rotating particles, the final spinner gel remains structurally isotropic, with no persistent chirality or odd mechanical responses. 
However, the elastic modulus is highly tunable through spinning activity, highlighting the role of active rotation in controlling gel mechanics. 
The key factor is network topology, where connectivity and loop formation govern rheology. 
While this work mainly reports a specific concentration ($\phi=0.4$), the observed trends hold across a range of $\phi$ from 0.3 to 0.5 (Supplementary Note 3), suggesting generic physics in spinner systems.
Preliminary results indicate that mixtures of counter-spinning particles can gel much more rapidly and form stiffer networks than unary spinner gels (Supplementary Note 5).
These findings contribute to the broader understanding of self-assembled, active gels, offering potential design principles for programmable soft matter and chiral metamaterials.


\section*{Methods}

\paratitle{LBM.}
We employ a coupled Lattice Boltzmann Method (LBM) and Discrete Element Method (DEM) framework to simulate active spinners in a Newtonian fluid.
The fluid dynamics are governed by the LBM with the Bhatnagar–Gross–Krook (BGK) single-relaxation-time approximation\,\cite{PhysRev.94.511}, in which the evolution of the distribution function $f_i$ follows:
\begin{equation}
f_i(\mathbf{r} + \mathbf{c}_i \Delta t, t + \Delta t) - f_i(\mathbf{r}, t) = - \frac{1}{\tau}[f_i(\mathbf{r}, t) - f_i^{\mathrm{eq}}(\mathbf{r}, t)],
\end{equation}
where $\mathbf{c}_i$ are the discrete lattice velocities (D3Q19), $\tau$ is the relaxation time, and $f_i^{\mathrm{eq}}$ is the local equilibrium distribution.


To account for the presence of solid particles, we adopt the partially saturated cell methods (PSM)\,\cite{IntJModPhysC.9.1189}, where each lattice node contains a local solid volume fraction $\varepsilon$.
The post-collision distribution is then given by:
\begin{equation}
f_i^{\text{new}} = (1 - \varepsilon) f_i^{\text{fluid}} + \varepsilon f_i^{\text{solid}},
\end{equation}
with $f_i^{\text{fluid}}$ and $f_i^{\text{solid}}$ computed from BGK collisions using the local fluid and solid velocities, respectively.
The superposition model, in which $f_i^{\text{solid}}$ is formed by linearly combining equilibrium distributions of the fluid and solid phases, is used to improve accuracy near curved or moving boundaries.


The fluid-solid interactions are then communicated to the DEM as hydrodynamic force (and torque), computed through momentum exchange with the surrounding fluid:
\begin{equation}
\mathbf{F}_\text{hydro} = \sum_i \mathbf{c}_i \left(f_i^{\text{solid}} - f_i^{\text{fluid}}\right),
\end{equation}
summed over lattice directions intersecting the particle boundary. 
Lubrication corrections\,\cite{PhysRevE.66.046708} are applied when interparticle gaps fall below the lattice size, compensating for under-resolved near-field interactions. 
This coupled DEM-LBM-PSM scheme captures both bulk and local hydrodynamics with high fidelity and efficiency, making it well-suited for studying the collective behavior of active systems.


\medskip
\paratitle{DEM.}
The DEM is implemented using LAMMPS\,\cite{CompPhysComm.271.108171}, where particle dynamics, including both translational and rotational motion, are governed by Newton's laws:
\begin{equation}
\begin{aligned}
m\diff{\mathbf{v}}{t} &= \mathbf{F}_\text{adh} + \mathbf{F}_\text{hydro}+ \mathbf{F}_\text{lub},
\\
I\diff{\mathbf{\omega}}{t} &= T_\text{adh} + T_\text{hydro}+ T_\text{lub}+ T_\text{a},\label{eq5}
\end{aligned}
\end{equation}
where $m$ and $I$ denote the mass and moment of inertia of a particle, respectively.
The hydrodynamic force $\mathbf{F}_\text{hydro}$ and torque $T_\text{hydro}$, as well as the lubrication corrections $\mathbf{F}_\text{lub}$ and $T_\text{lub}$, are computed from the LBM solver.
The active torque $T_{\text{a}}$, applied along the $-z$ direction, is constant and drives clockwise self-rotation of individual particles.


As illustrated in \figref{fig1}{b}, the adhesive interaction consists of a radial attraction and tangential constraints.
The attraction is modeled using a shifted Hookean spring with stiffness $k$, acting over a short range $\zeta_{\text{a}} = 0.01d$ beyond the particle surface.
This is equivalent to a truncated harmonic potential with a minimum energy of $U_{\text{adh}} = \frac{1}{2}k\zeta_{\text{a}}^2$.
Tangential constraints, including sliding and rolling resistance, are modeled by a modified Coulomb friction.
In particular, the tangential force is characterized by the same stiffness $k$ and a frictional coefficient $\mu=1$, expressed as
\begin{equation}
\mathbf{F}_\text{t} = -\min{\left[\mu (F_{\text{n}} + k\zeta_{\text{a}}), k\Delta_{\text{t}}\right]}\frac{\mathbf{v}_\text{t}}{||\mathbf{v}_\text{t}||},
\end{equation}
where $F_{\text{n}}$ refers to the normal contact force, $\Delta_{\text{t}}$ and $\mathbf{v}_\text{t}$ to the relative tangential displacement and velocity, respectively.
Frictional torque is computed analogously using rolling displacement and angular velocity.
See Ref.\,\cite{NatCommun.14.2773} for more details of adhesion models.

\medskip
\paratitle{Parameterization.}
Our system consists of spherical particles (of diameter $d$ and mass $m$) settling down to the bottom of a thin square box (of side lengths $L_{x, y} = 120d$ and height $h=3d$).
A Newtonian fluid (of viscosity $\eta$ and density $\rho$) is confined between the top and bottom solid no-slip walls in the $z$ directions.
While the hydrodynamics is solved in 3D, the particles are confined to the $x$-$y$ plane at $z = 0.5d$, forming a quasi-2D monolayer.
This setup mimics density-mismatched systems in experiments.
Through simulations, we confirm that such confinement yields results nearly identical to those obtained under moderate gravitation.
Importantly, this setup avoids complications arising from purely 2D hydrodynamics and provides a fundamental platform for chirality investigation.


The LBM is carried out with a lattice spacing of $\Delta x = 0.25d$ and a dimensionless relaxation time of $\tau = 0.68$. 
The consequent sound speed $c_{\text{s}}$ defines the smallest timescale in the simulation.
Hydrodynamics is updated every 100 DEM steps to ensure a balance between accuracy and computational efficiency.
Lubrication corrections are applied between particles to account for near-field hydrodynamic interactions unresolved by the lattice.
These include pairwise forces and torques arising from different modes of relative motion.
To prevent divergence, the lubrication is held constant below an inner cutoff of $10^{-3}d$, while an outer cutoff of $\Delta x$ ensures that corrections are only applied where lattice resolution fails. 
While we add lubrication corrections between particles, they are not applied to the particle-wall interactions, which, as we confirm, does not qualitatively change the results.


To ensure overdamped dynamics, we maintain the condition that the characteristic damping time, $m / 3\pi\eta d$, remains smaller than the interaction time scale $\sqrt{m/k}$.
Both timescales are set to be at least an order of magnitude larger than the sound-crossing time $d / c_{\text{s}}$.
The adhesive interaction is modeled using a Hookean spring with stiffness $k$ and interaction range $\zeta_{\text{a}} = 0.01d$, resulting in adhesion energy $U_{\text{adh}} = \frac{1}{2}k\zeta_{\text{a}}^2$.
To enforce strong bonding, we ensure that $U_{\text{adh}} \gg T_{\text{a}}$, where $T_{\text{a}}$ is the active torque applied to each particle to induce self-rotation.
The resulting spinning frequency $\omega$ defines a timescale $\omega^{-1}$, which is longer than all other dynamical timescales and thereby leads to quasi-static rotation (low Stokes number) and inertialess conditions ($\text{Re} < 1$).


Initial configurations are generated via Langevin dynamics of hard-sphere colloids with a temporarily enlarged diameter of $1.2d$ to prevent overlap.
The system is then evolved for a sufficiently long time to allow stable gelation.
For rheological measurements, the system is first relaxed and then sheared under a steady strain rate $\dot{\gamma}$, with $\dot{\gamma}^{-1}$ kept significantly larger than all intrinsic timescales to ensure quasi-static response.

\section*{Data availability}
The data that correlate with this study are available from the corresponding authors upon reasonable request.

\section*{Code availability}
The codes of the computer simulations are available from the corresponding authors upon reasonable request.

\section*{Acknowledgments}
We thank Zaiyi Shen and Zhiyuan Zhao for fruitful discussions.
This work was supported by National Natural Science Foundation of China (No.\,12404235) and Fundamental Research Funds for the Central Universities.

\section*{Author contributions}
Y.J. conceive the research. Y.J. and Y.C. carry out the simulation. All authors analyze the data. Y.J. and H.L. write the manuscript.

\section*{Competing interests}
The authors declare no competing interests.

\end{document}